\documentclass{optica-article}

\journal{opticajournal}
\articletype{Research Article}

\usepackage[T1]{fontenc}
\usepackage[utf8]{inputenc}
\usepackage{amsmath}
\usepackage{graphicx}
\usepackage{booktabs}
\usepackage{siunitx}
\usepackage{lineno}
\graphicspath{{figures/}}
\DeclareSIUnit{\dB}{dB}
\DeclareSIUnit{\bit}{bit}

\newcommand{\ket}[1]{\left|#1\right\rangle}
\newcommand{\bra}[1]{\left\langle#1\right|}

\begin{document}

\title{Design of an Electrically Tunable Microtoroid for Frequency Selection of Polarization-Entangled Photons}

\author{
Yichi Zhang\authormark{1},
Enqi Ke\authormark{1},
and Judith Su\authormark{1,2,*}
}

\address{\authormark{1}Wyant College of Optical Sciences, The University of Arizona, Tucson, Arizona 85721, USA}

\address{\authormark{2}Department of Biomedical Engineering, The University of Arizona, Tucson, Arizona 85721, USA}

\email{\authormark{*}judy@optics.arizona.edu}

\begin{abstract*}
Encoding quantum information into discrete optical frequencies, or ``frequency bins,'' uses different colors of light as additional information channels, allowing each photon to carry more information than polarization alone. However, combining conventional polarization-entangled photon sources with devices that manipulate these frequency channels requires a frequency selector that does not disturb the photons' fragile polarization entanglement. We present a computational design for an electrically tunable silica microtoroid that performs this function by selecting the desired frequency channels after the entangled photon pair has been generated. In the proposed architecture, the \(\SI{750}{\nano\meter}\) signal photon passes through the microtoroid, while its entangled \(\SI{880}{\nano\meter}\) partner bypasses the resonator and serves as a reference for the selected frequency channel. The principal challenge is resonator birefringence: because horizontally and vertically polarized light resonate at slightly different frequencies, the selected frequency can unintentionally reveal the photon's polarization state, weakening the quantum correlation between the photon pair. We solve this problem by adding a small lithium-niobate tuning element that is controlled with a single applied voltage. The voltage shifts the resonator so that it responds almost identically to horizontally and vertically polarized light, reducing the remaining mismatch to only 0.286 optical linewidths across all nine frequency channels. As a result, the photons remain strongly entangled after passing through the device, with a concurrence of \(C=0.969\), a Bell-state fidelity of \(F=0.981\), and a Bell parameter of \(S_{\max}=2.785\). If the relative timing between the different frequency channels is also controlled, the same device can generate a nine-channel polarization--frequency hyperentangled state, simultaneously encoding entanglement in both polarization and frequency, with an effective dimension of K=8.97. This computational design provides a compact, electrically tunable bridge between today's polarization-entangled photon sources and future high-capacity quantum photonic systems.
\end{abstract*}

\section{Introduction}

Photonic entanglement is a central resource for quantum communication, quantum networking, and optical quantum information processing.  Polarization-entangled photon-pair sources are well -established and widely used, but polarization provides only two orthogonal states per photon, limiting the amount of quantum information that can be encoded in this degree of freedom. Frequency-bin encoding treats different colors of light like separate information channels. A photon can travel in one of these channels or, because of quantum mechanics, in several of them at the same time.  When polarization and frequency are both entangled, the resulting hyperentangled state can carry more information per photon pair and can enable more efficient high-dimensional quantum protocols \cite{Kues2017,Imany2018,Lukens2017,Lu2023FrequencyBinReview}.

Integrated microresonators are powerful tools for frequency-bin quantum photonics.  Ring resonators have been widely used as photon-pair sources, filters, and frequency-bin processors \cite{Suo2015MicroringHyper,Xie2015Hyper,Rielander2018,Lu2023Hyper}.  In most of these demonstrations, the resonator participates directly in pair generation.  Here, we use an ultra-high-\(Q\) whispering gallery mode microtoroid resonator \cite{choi_impact_2022} as a frequency selector placed after an external source of polarization-entangled photon pairs (Bell pairs). Its role is to isolate a desired set of frequency bins while preserving the polarization entanglement of the photon pair. This role is not equivalent to ordinary spectral filtering. A useful selector must isolate the desired signal frequencies while ensuring that the selected frequency does not reveal the photon's polarization.

The problem arises because realistic resonators are birefringent.  If H- and V-polarized components of a Bell state are transmitted at slightly different resonance frequencies, the signal frequency becomes correlated with polarization.  The resonator has then partially measured the photon, and the polarization Bell state is degraded.  A Bell-compatible selector must therefore satisfy a stricter condition than high optical transmission: the H- and V-like resonance families must be aligned within the loaded linewidth across all accepted bins, and the access channel must not introduce large polarization imbalance or leakage.

We propose an electrically tunable silica microtoroid as a compact frequency selector that satisfies these requirements. A thin-film lithium-niobate (LN) tuning element with patterned electrodes is positioned adjacent to the outer rim of the microtoroid, where the whispering gallery mode evanescent field can be accessed. The applied voltage generates an electric field that tunes the resonator while avoiding direct contact with the optical mode. A single voltage shifts the H- and V-like mode families into near coincidence over the selected modes.  The signal photon passes through this tunable add--drop resonator channel, while the idler photon bypasses the resonator and identifies the matched energy-conserving bin.  If the source and reference arm preserve or program the relative phases among the accepted bin pairs, the selected output can be interpreted as a polarization--frequency hyperentangled state.

This architecture is intended to fill an intermediate role between static optical filters and fully programmable frequency-bin processors. Static bandpass filters, arrayed-waveguide gratings, and fixed microring resonators can define spectral channels, but they are not designed to preserve polarization entanglement as the primary performance criterion and cannot electrically compensate for resonator birefringence or drift. At the other extreme, pulse shapers and electro-optic frequency-bin processors can perform general frequency transformations, but they require substantially more active hardware than is needed for simple frequency selection \cite{Weiner2011PulseShaping,Lu2018EOBeamSplitter,Hu2021EOShifter}. The proposed microtoroid occupies the middle ground: a single applied voltage retunes all nine selected resonances while preserving polarization entanglement across the selected channels. It therefore provides a compact, electrically tunable frequency selector that allows conventional polarization-entangled photon sources to work with high-dimensional quantum systems that encode information using multiple frequency channels. The proposed architecture is based on the actively stabilized, ultra-high-Q silica microtoroid platform developed for the FLOWER sensing system \cite{su_accelerating_2026, hao_whispering_2024-1, kim_methotrexate_2024-4, gin_towards_2024, young_yang_agonist_2024, zhang_flower_2025, xu_free-space-coupled_nodate}, extending this established resonator technology from ultra-sensitive optical sensing to quantum-compatible frequency selection. 

Using this platform, we present a computational design study of a source that generates signal and idler photons at \SI{750}{\nano\meter} and \SI{880}{\nano\meter}, respectively. We ask whether a realistic microtoroid can select nine signal frequency channels within a \SI{10}{\nano\meter}-wide optical window while preserving the polarization entanglement of the photon pair. Our simulations show that the answer is yes. For a microtoroid with \(D=\SI{215.4}{\micro\meter}\) and \(a=b=\SI{4}{\micro\meter}\), numerical modeling predicts that a single applied voltage reduces the frequency mismatch between the horizontal and vertical polarization modes to just 0.286 optical linewidths while preserving strong Bell-state entanglement.

\section{Concept and operating principle}

\begin{figure}[t]
\centering
\includegraphics[width=\textwidth]{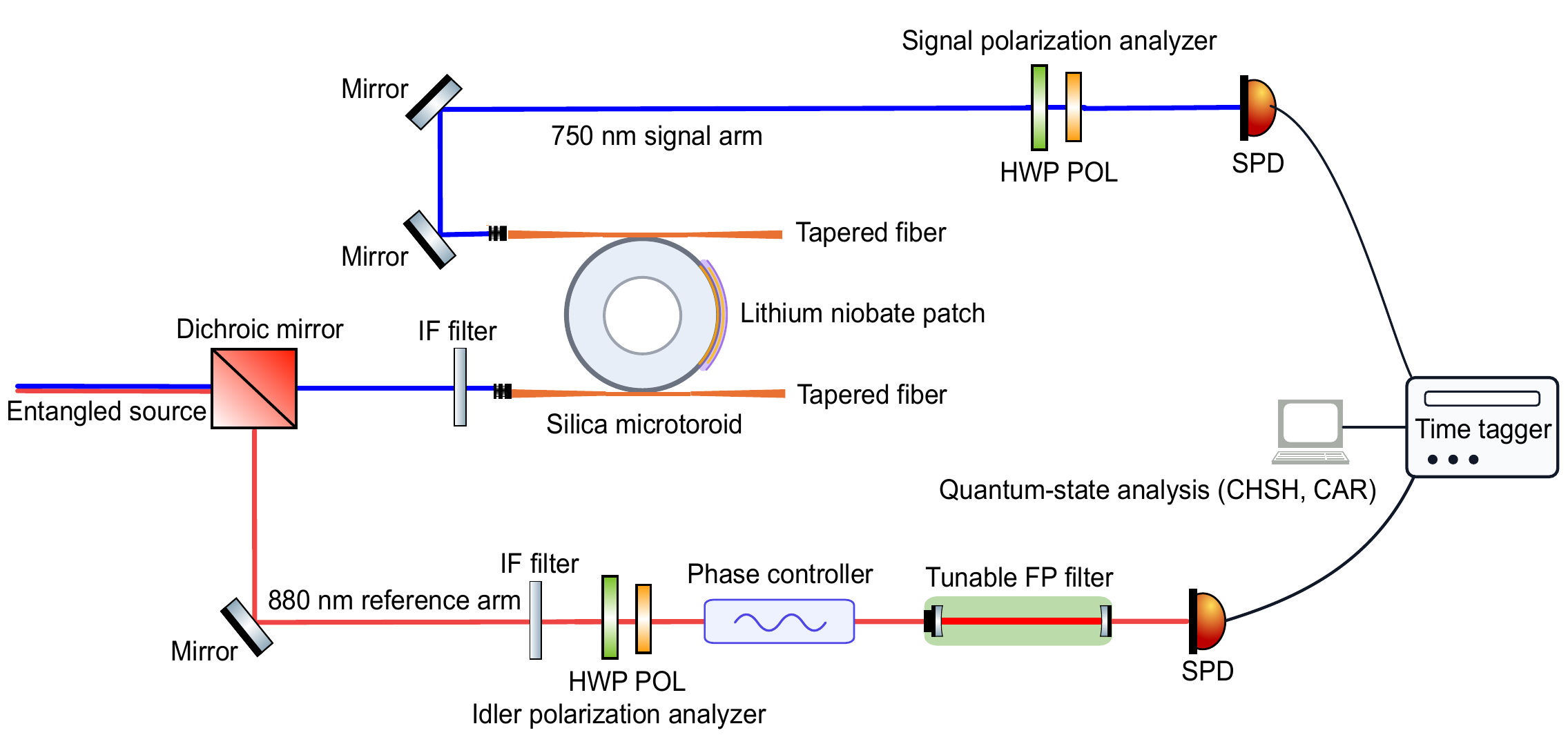}
\caption{\textbf{Experimental concept for polarization-preserving frequency selection of entangled photons.}
A conventional photon-pair source generates polarization-entangled signal--idler photon pairs distributed across multiple matched frequency channels. The \(\SI{750}{\nano\meter}\) signal photon is coupled into a voltage-gated silica microtoroid, which selects nine resonant frequency channels. A lithium-niobate (LN) tuning patch and electrodes apply an electric field that aligns the resonances for horizontally (H) and vertically (V) polarized light, preventing the selected frequency from revealing the photon's polarization. The entangled \(\SI{880}{\nano\meter}\) idler bypasses the microtoroid and, after polarization analysis and tunable filtering, identifies the corresponding energy-conserving frequency channel. Coincidences between the signal and idler photons are recorded with single-photon detectors (SPDs) and a time tagger to evaluate polarization entanglement and frequency-bin correlations.}
\label{fig:architecture}
\end{figure}

Figure~\ref{fig:architecture} illustrates the proposed architecture. An external photon-pair source generates polarization-entangled signal--idler photon pairs that are distributed across several matched frequency channels. The signal photon passes through the electrically tunable microtoroid, while the idler photon bypasses the resonator and is used to identify the corresponding frequency channel and characterize the output state. An idealized source state is\cite{Kues2017,Xie2015Hyper,Lu2023FrequencyBinReview}

\begin{equation}
\ket{\Psi_\mathrm{src}}=
\sum_k c_k e^{i\alpha_k}\ket{k_s,k_i}\otimes
\frac{\ket{H_sH_i}+e^{i\phi_0}\ket{V_sV_i}}{\sqrt{2}},
\qquad
\sum_k |c_k|^2=1 .
\label{eq:source_state}
\end{equation}

Here, \(k\) labels an energy-conserving signal--idler frequency-channel pair, \(c_k\) is the probability amplitude for that pair, and \(\alpha_k\) is its relative phase. The microtoroid acts only on the signal photon; the idler remains unchanged and serves as a reference for the selected frequency channel.

A coincidence measurement verifies the one-to-one correspondence between the selected signal and idler frequency channels\cite{Rielander2018,Cheng2023SinglyFilteredBFC,Lu2023FrequencyBinReview},

\begin{equation}
P_{mn}=
\mathrm{Tr}\!\left[
\rho\,\Pi^{(s)}_m\Pi^{(i)}_n
\right]
\simeq q_m\delta_{n,\pi(m)} ,
\label{eq:pair_correlation}
\end{equation}

where \(P_{mn}\) is the probability of detecting the signal photon in channel \(m\) and the idler photon in channel \(n\), \(\rho\) is the two-photon density matrix, \(\Pi^{(s)}_m\) and \(\Pi^{(i)}_n\) are measurement operators  that select the corresponding signal and idler frequency channels, \(q_m\) is the probability of signal channel \(m\), and \(\pi(m)\) identifies the idler channel paired with signal channel \(m\). This one-to-one frequency correlation is required, but by itself does not produce a high-dimensional frequency-bin quantum state. The different frequency-channel pairs must also maintain fixed relative phases so that they form a coherent superposition. In the proposed experiment, these phases can be stabilized using a reference-arm delay or fiber stretcher and, if necessary, adjusted using pulse shaping or electro-optic phase control. If the phases \(\alpha_k\) become random, the output remains a collection of independent polarization-entangled frequency channels rather than a coherent frequency-bin superposition.

The microtoroid acts independently on each selected frequency channel. Before including the small off-diagonal polarization-mixing (Jones leakage) terms, its transfer matrix for frequency channel \(k\) is\cite{Yariv2000,Bogaerts2012Microring}

\begin{equation}
T_k(V)=
\begin{pmatrix}
t_{H,k}(V)e^{i\theta_{H,k}(V)} & 0\\
0 & t_{V,k}(V)e^{i\theta_{V,k}(V)}
\end{pmatrix},
\label{eq:selector_transfer}
\end{equation}

where \(t_{H,k}\) and \(t_{V,k}\) are the transmission amplitudes for horizontally (H) and vertically (V) polarized light, respectively, and \(\theta_{H,k}\) and \(\theta_{V,k}\) are the corresponding phase shifts introduced by the resonator.

To preserve the polarization-entangled state, the resonator must respond almost identically to horizontally (H) and vertically (V) polarized light. Otherwise, the selected frequency would reveal the photon's polarization, reducing the entanglement. The central design goal is therefore to align the H- and V-like resonances as closely as possible across all selected frequency channels.

For the two WGM polarization families, the resonance frequencies are approximated by\cite{vahala2003optical,Oxborrow2007,Guarino2007}

\begin{align}
\nu_{p,k}(V)
&\simeq
\nu_{p,0}+kD_{1,p}+\frac{k^2}{2}D_{2,p}+g_pV,\\
\delta_k(V)
&\equiv \nu_{H,k}(V)-\nu_{V,k}(V)
\simeq
\Delta\nu_0+k\Delta D_1+\frac{k^2}{2}\Delta D_2+\Delta g\,V .
\label{eq:one_voltage_dispersion}
\end{align}

Here, \(\nu_{p,k}\) is the resonance frequency of polarization family \(p\) (H or V) and mode \(k\), \(\nu_{p,0}\) is the reference resonance frequency, \(D_{1,p}\) is the free spectral range, \(D_{2,p}\) describes the dispersion of the mode family, \(g_p\) is the electro-optic tuning coefficient, and \(V\) is the applied voltage. The quantity \(\delta_k(V)\) is the frequency separation between the H- and V-like resonances for mode \(k\).

Applying a single voltage removes the overall H/V frequency offset but cannot compensate arbitrary differences in dispersion between the two mode families. Consequently, the microtoroid geometry must be designed so that the H- and V-like resonance spectra remain nearly parallel over the selected frequency channels. The 215 micron diameter microtoroid satisfies this requirement, allowing a single lithium-niobate tuning voltage to align all nine selected resonances.

\section{Microtoroid design and electro-optic alignment}

The proposed device is based on a silica microtoroid with a major diameter of \(D=\SI{215.4}{\micro\meter}\) and minor radii of \(a=b=\SI{4}{\micro\meter}\). This geometry was selected because it supports two nearly parallel whispering gallery mode (WGM) polarization families that can be aligned with a single applied voltage. Axisymmetric COMSOL eigenmode simulations were used to calculate the optical resonances. The selected H-like and V-like mode families consist of modes H1311--H1319 and V1306--V1314, respectively, centered near \(\SI{749.66}{\nano\meter}\).

To compare the frequency separation between the H and V resonances, we express the residual mismatch in units of the loaded optical linewidth. Assuming a loaded quality factor of \(Q_L=10^6\), the linewidth of resonance \(k\) is\cite{vahala2003optical,Bogaerts2012Microring}

\begin{equation}
\Delta\nu_{\mathrm{LW},k}=\frac{c}{\lambda_k Q_L},
\label{eq:linewidth}
\end{equation}

where \(c\) is the speed of light and \(\lambda_k\) is the resonance wavelength. Expressing the H/V mismatch in linewidth units provides a direct measure of whether the resonances are sufficiently aligned to preserve polarization entanglement.

The numerical accuracy of the simulations was verified using coarse, medium, fine, and extra-fine computational meshes. The production fine mesh contains \(1.09\times10^5\) elements and \(5.47\times10^4\) vertices, with a maximum element size of \(\SI{75}{\nano\meter}\) near the resonator rim. Compared with the extra-fine mesh, the fine mesh changes the maximum H/V resonance splitting by only \(\SI{16.1}{\mega\hertz}\), which is much smaller than the loaded optical linewidth relevant for frequency selection. The fine mesh therefore provides sufficient numerical accuracy and is used for all reported optical-field and quantum-state calculations.
\begin{figure}[t]
\centering
\includegraphics[width=\textwidth]{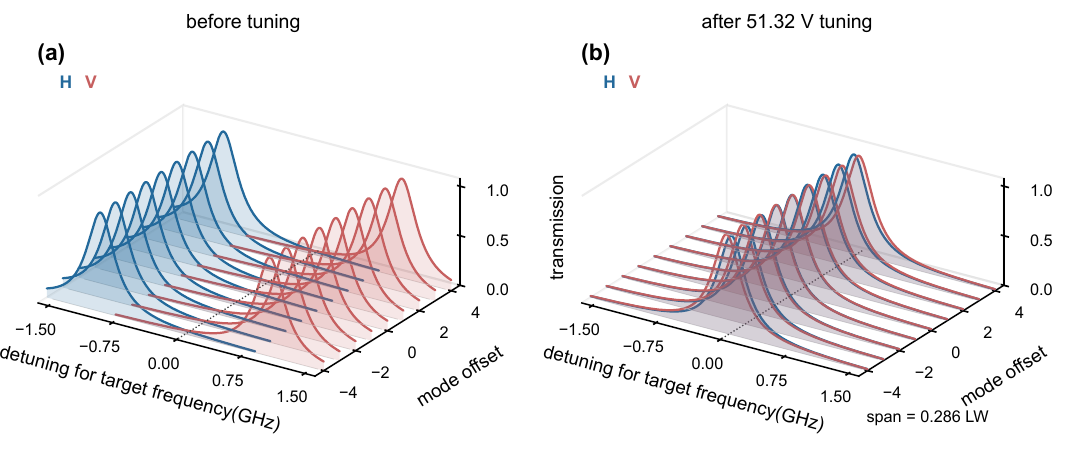}
\caption{\textbf{Electro-optic alignment of the horizontal (H) and vertical (V) resonance families.} (a) Before tuning, the H- and V-like resonances are separated by approximately \(\SI{2}{\giga\hertz}\). As a result, the transmitted frequency depends on the photon's polarization, which can degrade polarization entanglement. (b) Applying a \(\SI{51.32}{\volt}\) bias to the lithium-niobate (LN) actuator shifts the resonances into near coincidence across all nine selected modes. The remaining H/V frequency mismatch is only 0.286 loaded linewidths, allowing frequency selection while preserving the polarization-entangled state.}
\label{fig:line_tuning}
\end{figure}

Figure~\ref{fig:line_tuning} shows the key alignment result.  Before tuning, the H- and V-like resonances are separated by approximately \(\SI{2}{\giga\hertz}\).  A quarter-rim LN/electrode actuator supplies a differential electro-optic correction of \(\SI{2.035}{\giga\hertz}\) at \(\SI{51.32}{\volt}\), leaving a residual H/V span of 0.286 loaded linewidths and a maximum absolute residual of 0.143 linewidths across the nine modes.  This residual is small enough that the selected frequency bins do not strongly distinguish the two polarization components.

The applied voltage shifts the resonator frequencies through the electro-optic (Pockels) effect in the lithium-niobate (LN) tuning layer. This frequency shift is calculated using first-order perturbation theory\cite{Johnson2002Perturbation,Guarino2007,WeisGaylord1985},

\begin{equation}
\frac{\Delta \omega_p}{\omega_p}
=
-\frac{1}{2}
\frac{
\displaystyle
\int
\mathbf{E}_p^{*}(\mathbf{r})
\cdot
\Delta\boldsymbol{\varepsilon}_p(\mathbf{r},V)
\cdot
\mathbf{E}_p(\mathbf{r})
\, d^{3}r
}{
\displaystyle
\int
\mathbf{E}_p^{*}(\mathbf{r})
\cdot
\boldsymbol{\varepsilon}_p(\mathbf{r})
\cdot
\mathbf{E}_p(\mathbf{r})
\, d^{3}r
},
\label{eq:eo_perturbation}
\end{equation}

where \(\Delta\boldsymbol{\varepsilon}_p\) is the voltage-induced change in the dielectric permittivity tensor produced by the Pockels effect in the lithium-niobate tuning layer, \(\boldsymbol{\varepsilon}_p\) is the unperturbed dielectric permittivity tensor, and \(\mathbf{E}_p\) is the optical electric-field distribution of the corresponding whispering-gallery mode. The frequency shift depends on the overlap between the applied electric field and the optical whispering-gallery mode.

The proposed tuning structure consists of a \(\SI{700}{\nano\meter}\)-thick LN film positioned \(\SI{20}{\nano\meter}\) from the microtoroid rim, together with two \(\SI{50}{\nano\meter}\) electrodes that generate the tuning electric field. Figure 3 illustrates this geometry along with the simulated electrostatic and optical fields. For the central resonance pair, the calculated electro-optic tuning coefficients, defined as the resonance frequency shift per unit applied voltage\cite{Guarino2007,WeisGaylord1985}, are

\[
\gamma_H=-\SI{49.37}{\mega\hertz\per\volt},\qquad
\gamma_V=-\SI{9.71}{\mega\hertz\per\volt},
\]

for the H- and V-like resonance families, respectively, 
giving a differential tuning coefficient of \(-\SI{39.65}{\mega\hertz\per\volt}\). Because the optical field distribution changes slightly from one frequency channel to the next, the tuning coefficient is calculated separately for each selected mode before determining the single operating voltage that best aligns all nine H and V resonances.
\begin{figure}[t]
\centering
\includegraphics[width=\textwidth]{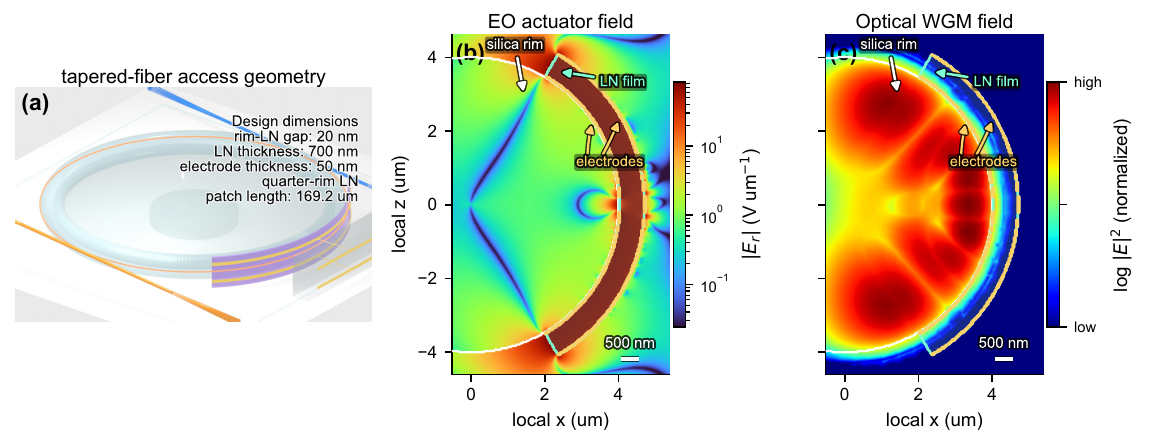}

\caption{\textbf{Microtoroid geometry and electro-optic tuning concept.} (a) Schematic of the proposed device. Tapered input and output fibers provide add--drop coupling to the silica microtoroid, while a quarter-rim lithium-niobate (LN) patch and electrodes apply the electro-optic tuning field. (b) COMSOL simulation of the electric-field magnitude produced by the LN actuator at the operating voltage. (c) Simulated optical whispering gallery mode (WGM) intensity in the same cross-section. The overlap between the electric field and the optical mode determines how efficiently the applied voltage shifts the resonator frequency through the electro-optic effect, as described by ~Eq.~\eqref{eq:eo_perturbation}.}
\label{fig:taper_access}
\end{figure}

In an experiment, resonance tracking, thermal stabilization, and active voltage feedback would be required to maintain the alignment point.  The present calculation should therefore be read as a design target: it identifies the resonator geometry, actuator scale, and H/V tolerance needed for a Bell-compatible selector.

\section{Coupled-mode/Jones model and Bell-state preservation}

Aligning the H- and V-like resonances is necessary but not sufficient to preserve the polarization-entangled state. The resonator must also couple light into and out of the H and V WGM mode families with nearly equal efficiency while minimizing polarization mixing. To estimate these coupling strengths, we use an isolated-mode coupled-mode model\cite{HausHuang1991,Yariv2000,Pfeiffer2017CouplingIdeality},

\begin{equation}
\eta_{p,k}^{(r)}=\eta_0^{(r)}
\frac{
\left(F_{\mathrm{surf},p,k}/V_{\mathrm{mode},p,k}\right)
\mathrm{sinc}^2(\Delta\beta_{p,k}L/2)
}{
\left\langle
\left(F_{\mathrm{surf}}/V_{\mathrm{mode}}\right)
\mathrm{sinc}^2(\Delta\beta L/2)
\right\rangle
},
\label{eq:relative_cmt}
\end{equation}

where \(p\in\{H,V\}\) denotes the polarization, \(r\) denotes the input or drop coupler, \(F_{\mathrm{surf}}\) is the optical field at the resonator surface, \(V_{\mathrm{mode}}\) is the optical mode volume, and \(\Delta\beta\) is the phase mismatch over the interaction length \(L\). The parameter \(\eta_0^{(r)}\) sets the overall coupling efficiency and is used only when estimating photon count rates. The polarization-entanglement metrics depend primarily on the relative balance between the H and V coupling strengths and on any polarization leakage.

The effect of the resonator on each selected frequency channel is described by the Jones matrix\cite{Shadbolt2012,Carolan2015Universal}

\begin{equation}
J_k=
\begin{pmatrix}
\sqrt{\eta_{H,k}} & \ell_{HV,k}\\
\ell_{VH,k} & \sqrt{\eta_{V,k}}
\end{pmatrix},
\label{eq:jones_cmt}
\end{equation}

where \(\eta_{H,k}\) and \(\eta_{V,k}\) are the transmission efficiencies for the horizontally (H) and vertically (V) polarized components of frequency channel \(k\), respectively. The off-diagonal terms, \(\ell_{HV,k}\) and \(\ell_{VH,k}\), describe unwanted polarization leakage, in which H-polarized light is converted to V polarization or vice versa. Applying \(J_k\) to the signal photon in Eq.~\eqref{eq:source_state} predicts the polarization state after frequency selection. The frequency information is then removed by tracing over the selected frequency channels, yielding the output polarization density matrix. The reconstructed polarization state is then used to calculate the concurrence, Bell-state fidelity, and the maximum Clauser--Horne--Shimony--Holt (CHSH) Bell parameter, which together quantify the strength and quality of the remaining polarization entanglement.
\begin{figure}[t]
\centering
\includegraphics[width=\textwidth]{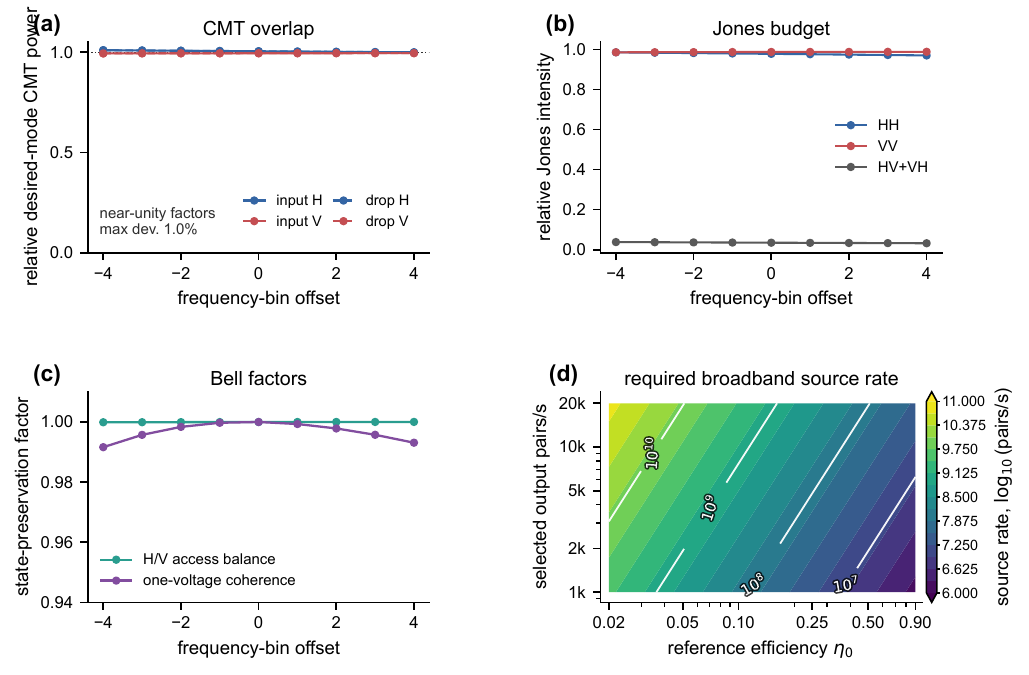}
\caption{\textbf{Coupled-mode and polarization analysis of the proposed frequency selector.} (a) Simulated input and drop coupling strengths for the horizontal (H) and vertical (V) whispering gallery mode families across the nine selected frequency channels. (b) Jones-matrix analysis showing that light remains almost entirely in its original polarization (HH and VV), with only small polarization-mixing (HV+VH) terms. (c) Frequency-channel dependence of the H/V coupling balance and the residual coherence after applying a single tuning voltage, demonstrating nearly uniform performance across all nine selected channels. (d) Estimated broadband photon-pair source rate required to achieve a target detected pair rate as a function of the reference coupling efficiency \(\eta_0\).}
\label{fig:cmt_jones}
\end{figure}

The calculated coupling strengths remain well balanced across the nine selected frequency channels. Figure 4a–c shows the simulated coupling strengths, Jones budget, and Bell-compatibility factors across the nine channels, while Fig. 4d estimates the required source brightness. The H-like modes vary from 0.999 to 1.020, while the V-like modes vary from 0.989 to 0.991, giving H/V coupling ratios between 1.008 and 1.032. After including the remaining H/V phase differences and the small amount of polarization leakage predicted by the Jones model, the output state retains strong polarization entanglement\cite{Wootters1998,Horodecki1995,CHSH1969},

\[
C=0.969,\qquad
F=0.981,\qquad
S_{\max}=2.785 .
\]

These results show that the proposed frequency selector preserves polarization entanglement even in the presence of small imperfections. We then asked how much worse the device could become before it no longer preserved a Bell state. To answer this, we artificially increased the polarization leakage in the model. The Bell inequality remained violated until the leakage amplitudes were increased by approximately \(8.33\times\), showing that the design has a substantial safety margin and is not operating close to its performance limit.

Next, we ask how well the nine selected frequency channels are used. Ideally, photons should be distributed evenly among all nine channels so that each contributes to the quantum state. The probability \(q_k\) of occupying each selected frequency channel is therefore used to calculate the effective dimension \(K\), which estimates how many channels are effectively used, and the Shannon entropy \(H\), which measures how evenly the photons are distributed,

After selecting the nine frequency channels, we evaluate how effectively they are used. A more uniform distribution allows more of the available frequency channels to contribute, increasing the amount of quantum information that can be encoded. The probability \(q_k\) of occupying each selected frequency channel determines the effective dimension \(K\) and the Shannon entropy \(H\),
After selecting the nine frequency channels, we ask two questions: How many of the channels are used, and how evenly are the photons distributed among them? These properties are quantified by the effective dimension \(K\), which estimates the number of frequency channels that contribute significantly, and the Shannon entropy \(H\), which measures how uniformly the photons are distributed among those channels. Both quantities are calculated from the probability \(q_k\) of occupying frequency channel \(k\)\cite{Shannon1948,LawEberly2004,Lu2023FrequencyBinReview},

\begin{equation}
K=\frac{1}{\sum_k q_k^2},
\qquad
H=-\sum_k q_k\log_2 q_k .
\label{eq:dimension}
\end{equation}

For the D215 microtoroid, the selected channels give \(K=8.97\) and \(H=3.17\) bits, both very close to the ideal values for nine equally populated frequency channels. The corresponding signal--idler frequency pairs are identified experimentally using the idler analyzer described by Eq.~\eqref{eq:pair_correlation}. In a practical implementation, the analyzer is tuned sequentially across the nine idler frequency channels, and the complete signal--idler correlation map is reconstructed from the measured coincidence counts.

If the relative phases between the selected frequency channels are stabilized or actively controlled, the photons form a coherent quantum state spanning all nine frequency channels. This state is described by the reduced frequency-bin density matrix\cite{Kues2017,Imany2018,Lukens2017,Lu2023FrequencyBinReview},

\begin{equation}
\rho_\mathcal{S}(V)
=
\frac{1}{P_\mathcal{S}}
\sum_{k,l\in\mathcal{S}}
c_k c_l^\ast
e^{i[(\alpha_k+\bar{\theta}_k)-(\alpha_l+\bar{\theta}_l)]}
\left\langle\chi_l(V)\middle|\chi_k(V)\right\rangle
\ket{k_s,k_i}\bra{l_s,l_i}.
\label{eq:frequency_density}
\end{equation}

\begin{figure}[t]
\centering
\includegraphics[width=\textwidth]{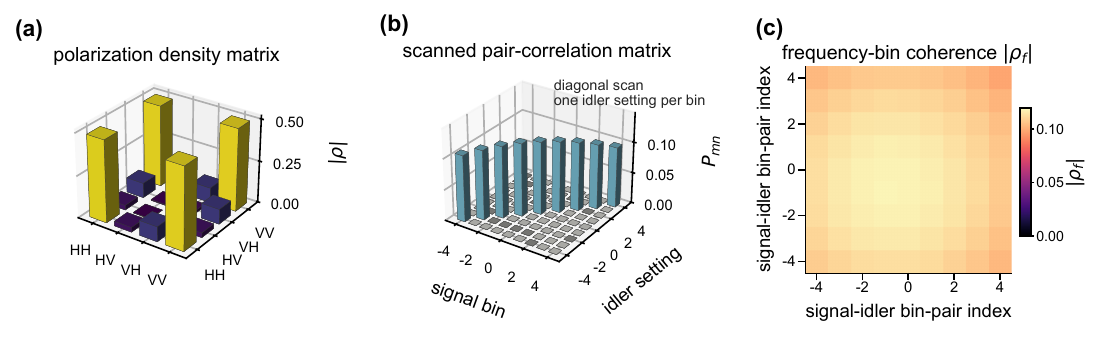}
\caption{\textbf{Verification of polarization entanglement and coherence between the selected frequency channels} (a) Polarization density matrix obtained after averaging over the selected frequency channels. The strong diagonal (HH and VV) and coherent off-diagonal terms indicate that the polarization entanglement is preserved after the photons pass through the microtoroid.  (b) Simulated signal response for each selected idler frequency channel. Each idler channel produces a single corresponding signal resonance, confirming the one-to-one frequency pairing established by energy conservation.
(c) Reduced frequency-bin density matrix after phase correction. The nearly uniform diagonal elements show that all nine selected frequency channels are occupied, while the strong off-diagonal elements show that the channels maintain fixed phase relationships rather than behaving independently. This preserved coherence is required to support high-dimensional frequency-bin quantum states.}
\label{fig:frequency_bin_analysis}
\end{figure}

Here, \(\mathcal{S}\) denotes the nine selected frequency channels, \(P_\mathcal{S}\) is the total probability of selecting those channels, \(c_k\) and \(\alpha_k\) are the source-defined amplitudes and phases introduced in Eq.~\eqref{eq:source_state}, \(\bar{\theta}_k\) is the phase added by the microtoroid, and \(\langle\chi_l(V)|\chi_k(V)\rangle\) accounts for the overlap of the corresponding polarization states after frequency selection.
The off-diagonal elements of this matrix describe the phase relationships between different frequency channels. They remain large only when the relative phases are stable or have been actively corrected; pair correlations alone do not guarantee this coherence. When the relative phases are stabilized and adjusted to be equal, the selected frequency state is nearly identical to an ideal coherent nine-channel state, with a purity of 0.99997, a coherent off-diagonal fraction of 0.99998, and a fidelity of \(F_9=0.99916\) to a uniformly populated nine-channel entangled state. The same device output can therefore be interpreted in two ways. Without phase stabilization, it acts as a polarization-entangled photon source with nine distinguishable frequency channels. With phase stabilization, the same output supports a high-dimensional polarization--frequency hyperentangled state. Figure~\ref{fig:frequency_bin_analysis} shows that the proposed frequency selector preserves the key properties of the quantum state. Panel (a) verifies that polarization entanglement is maintained, panel (b) confirms the expected one-to-one pairing between the selected signal and idler frequency channels, and panel (c) shows that coherence between the selected frequency channels can also be preserved when their relative phases are stabilized.

The previous sections evaluate the quality of the selected quantum state. We now estimate the photon rates expected in a practical experiment. These calculations are intended to determine whether the proposed selector can operate with realistic photon sources and detectors.

The overall signal transmission through the selector is\cite{Yariv2000,HausHuang1991,Pfeiffer2017CouplingIdeality}

\begin{equation}
T_\mathcal{S}
=
\eta_{0,\mathrm{in}}\eta_{0,\mathrm{drop}}
\sum_{k\in\mathcal{S}}p_k
\frac{
g_{H,k}^{(\mathrm{in})}g_{H,k}^{(\mathrm{drop})}
+
g_{V,k}^{(\mathrm{in})}g_{V,k}^{(\mathrm{drop})}
}{2},
\label{eq:throughput}
\end{equation}

where \(\eta_0\) is the measured coupling efficiency between the optical fibers and the microtoroid, \(p_k\) is the probability that a photon occupies frequency channel \(k\), and \(g_{p,k}^{(r)}\) are the relative coupling factors calculated from the coupled-mode/Jones model. A source prepared directly in the selected nine frequency channels has \(p_9=1\). In contrast, a broadband source distributes photons over a much wider spectrum, so only a small fraction (\(p_9=1.52\times10^{-3}\)) falls within the nine selected channels.

The corresponding idler frequency channel is identified using a scanned Fabry--Perot (FP) analyzer. Because both the analyzer and the idler frequency channels have finite linewidths, only part of the light in each channel is transmitted. Assuming Lorentzian transmission profiles for both\cite{SalehTeich2007,Pruessner2007IntegratedFP,ThorlabsFPQFA8}, the transmission of a single selected frequency channel is approximated by

\begin{equation}
T_\mathrm{bin}\simeq
T_\mathrm{peak}
\frac{\Delta\nu_\mathrm{FP}}
{\Delta\nu_\mathrm{bin}+\Delta\nu_\mathrm{FP}},
\label{eq:fp_bin_overlap}
\end{equation}

where \(T_\mathrm{peak}\) is the peak transmission of the analyzer, \(\Delta\nu_\mathrm{FP}\) is the Fabry--Perot linewidth, and \(\Delta\nu_\mathrm{bin}\) is the linewidth of the selected idler frequency channel. For the calculations presented here, we assume \(T_\mathrm{peak}=0.8\). A passband-matched analyzer with \(Q_\mathrm{FP}=5\times10^5\) gives a calculated single-channel transmission of \(T_\mathrm{bin}=0.504\).

Using this transmission together with the coupling efficiencies, we estimate the photon-pair source brightness required for the proposed experiment. Selected output-pair rates of \(2\times10^3\) to \(2\times10^4~\mathrm{s^{-1}}\) require prepared source rates of \(3.97\times10^5\) to \(3.97\times10^6~\mathrm{s^{-1}}\) for \(\eta_0=0.10\), or \(1.59\times10^4\) to \(1.59\times10^5~\mathrm{s^{-1}}\) for \(\eta_0=0.50\). Broadband photon sources require proportionally higher generation rates because only a small fraction of the generated photon pairs fall within the nine selected frequency channels.
Finally, we estimate the measured coincidence rates after including realistic detector losses and background counts. The selected photon-pair rate, the detected true coincidence rate, and the accidental coincidence rate are given by\cite{Eisaman2011,Hadfield2009}

\begin{align}
R_\mathrm{sel}&=R_\mathrm{tot}p_9T_\mathcal{S}T_\mathrm{bin},\\
C_\mathrm{true}&=R_\mathrm{sel}\xi_s\xi_i,\\
C_\mathrm{acc}&=S_sS_i\tau_c,\qquad
\mathrm{CAR}=\frac{C_\mathrm{true}}{C_\mathrm{acc}} .
\label{eq:car_noise}
\end{align}

Here, \(R_\mathrm{tot}\) is the total photon-pair generation rate, \(R_\mathrm{sel}\) is the rate after frequency selection, \(\xi_s\) and \(\xi_i\) are the signal and idler detector efficiencies, \(S_s\) and \(S_i\) are the total single-count rates (including dark counts and stray light), and \(\tau_c\) is the coincidence time window. The coincidence-to-accidental ratio (CAR) measures the number of true photon-pair coincidences relative to accidental coincidences caused by detector noise and background events.

In a real experiment, some detected photon pairs are true entangled pairs, while others are accidental coincidences caused by detector noise or stray background light. We model the measured quantum state as a combination of these two contributions\cite{Werner1989,Horodecki1995},

\begin{equation}
\rho_\mathrm{obs}
=
\frac{\mathrm{CAR}}{\mathrm{CAR}+1}\rho_\mathrm{corr}
+
\frac{1}{\mathrm{CAR}+1}\frac{\mathbf{I}_4}{4}.
\label{eq:car_density}
\end{equation}

where \(\rho_\mathrm{corr}\) is the ideal polarization-entangled state and \(\mathbf{I}_4/4\) is the maximally mixed polarization state, which models accidental coincidences as an uncorrelated white-noise background, and the coincidence-to-accidental ratio (CAR) determines the relative contribution of the two.

\begin{figure}[t]
\centering
\includegraphics[width=\textwidth]{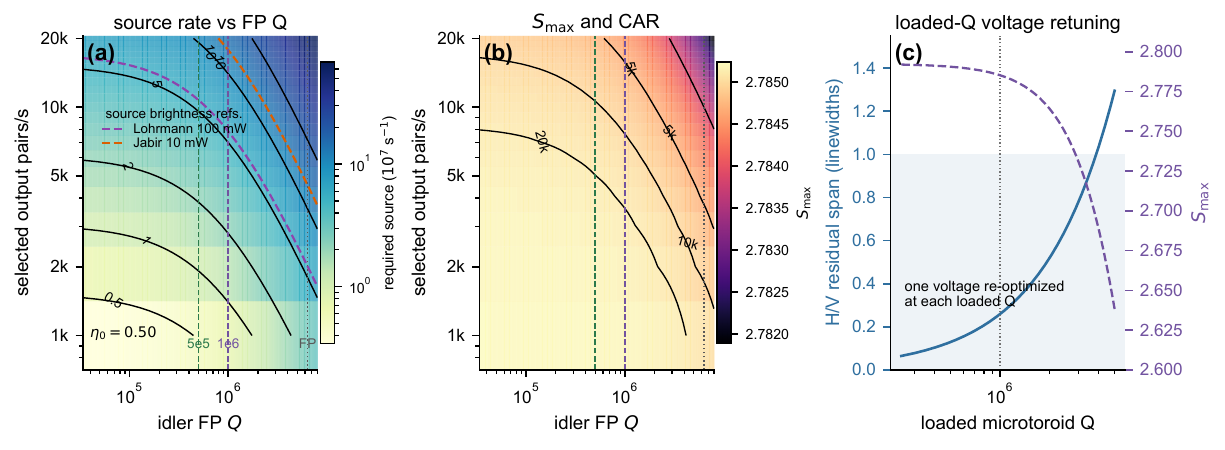}
\caption{\textbf{Predicted experimental performance of the proposed frequency selector.} (a) Photon-pair source rate required to achieve a desired detected signal--idler count rate after frequency selection. Better idler Fabry--P\'erot (FP) filter performance reduces the source brightness needed. Dashed source-brightness reference contours are estimated from published PPKTP photon-pair sources, using \(0.56\times10^6~\mathrm{pairs\,s^{-1}\,mW^{-1}}\) at \(\SI{100}{\milli\watt}\) from Lohrmann \emph{et al.} and \(0.41\times10^6~\mathrm{s^{-1}\,mW^{-1}\,nm^{-1}}\) over \(\SI{31}{\nano\meter}\) at \(\SI{10}{\milli\watt}\) from Jabir and Samanta \cite{Lohrmann2020BroadbandPPKTP,Jabir2017HighBrightness}. The vertical guide lines mark representative idler-FP filter choices, including \(Q_\mathrm{FP}=5\times10^5\) and \(10^6\), and the corresponding sub-\(\SI{100}{\mega\hertz}\)-class passband scale. (b) Predicted Bell parameter \(S_{\max}\) after including realistic detector imperfections, including finite detector efficiency, background counts, and accidental coincidences. The predicted Bell parameter remains well above the classical limit over the practical operating range, indicating that strong polarization entanglement is preserved. (c) Effect of the loaded microtoroid quality factor (\(Q\)) after re-optimizing the lithium-niobate (LN) tuning voltage. As \(Q\) increases, the resonances become narrower, making the H/V alignment more sensitive. The residual H/V frequency mismatch therefore increases, eventually reducing the optimized Bell parameter \(S_{\max}\).}

\label{fig:channel_quality}
\end{figure}

For a coincidence window of \(\tau_c=\SI{1}{\nano\second}\), detector efficiencies of \(\xi_s=0.455\) and \(\xi_i=0.420\), and dark-plus-stray count rates of \(D_s=D_i=200~\mathrm{s^{-1}}\), the calculated coincidence-to-accidental ratio is sufficiently high that the predicted Bell parameter remains well above the classical limit of \(S=2\). Figure~\ref{fig:channel_quality} summarizes the expected experimental performance. Panel (a) shows how bright the photon-pair source must be, panel (b) shows how detector imperfections affect the measured entanglement, and panel (c) shows how the resonator quality factor affects the alignment of the H- and V-like resonances.These calculations are intended as feasibility estimates; a specific experiment would use the measured source spectrum, coupling efficiencies, thermal stability, and detector characteristics of the experimental system.
\section{Discussion}

The central result of this work is that a resonator placed after a photon-pair source should be evaluated not only by its optical transmission, but also by its ability to preserve the quantum state. A practical frequency selector must isolate the desired frequency channels without allowing the transmitted frequency to reveal the photon's polarization. For the proposed D215 microtoroid, a single electro-optic tuning voltage aligns the horizontal- and vertical-like resonance families across nine selected frequency channels while maintaining strong Bell-state performance.

This role differs from that of conventional optical filters. Static filters can isolate a spectral band but cannot compensate resonator birefringence or adapt to thermal or mechanical drift. At the opposite extreme, pulse shapers and electro-optic frequency-bin processors provide much greater flexibility but require substantially more complex hardware. Instead, the proposed device provides a compact, electrically tunable frequency selector that allows conventional polarization-entangled photon sources to be used with quantum systems that encode information across multiple frequency channels.

The key design challenge is not simply achieving polarization-independent filtering at a single wavelength, but demonstrating that a compact high-\(Q\) resonator can preserve polarization indistinguishability across multiple frequency channels using only a single electrical control parameter. Because the H- and V-like whispering gallery mode families have slightly different free spectral ranges and dispersions, their frequency separation changes from one resonance to the next. A single applied voltage can remove only the common H/V offset; the remaining differential dispersion must therefore be sufficiently small by design. The D215 microtoroid satisfies this condition, allowing all nine selected resonances to be aligned simultaneously while providing active electrical tuning to compensate for source or resonator drift.

The present work is a computational design study rather than an experimental demonstration. Practical implementation will require stable operation of a high-\(Q\) microresonator, low-loss integration of the lithium-niobate tuning element, and experimental characterization of the source spectrum, frequency-channel phases, and coupling efficiencies. At \(Q_L=10^6\), the loaded linewidth is approximately \(\SI{400}{\mega\hertz}\), whereas the silica thermo-optic shift is several gigahertz per kelvin. Maintaining the required resonance alignment will therefore require active resonance tracking together with thermal and voltage stabilization. A practical proof-of-concept could position the thin-film lithium-niobate tuning element adjacent to the microtoroid using a nanopositioner, with permanent heterogeneous integration representing a subsequent engineering step.

The interpretation of the output as a polarization--frequency hyperentangled state also depends on maintaining coherent phase relationships among the selected frequency channels. Although the proposed architecture provides a practical route to achieving this condition, an experimental realization would require frequency-bin interference measurements or quantum-state tomography to verify the necessary coherence. Until such measurements are performed, the device should be regarded as a Bell-compatible frequency selector rather than a demonstrated hyperentanglement source.

Finally, the broader scaling analysis presented in the Supplementary Information indicates that the same design principles extend to larger odd-dimensional frequency-channel windows. The proposed architecture is therefore not limited to the nine-channel example studied here and may provide a scalable interface between conventional polarization-entangled photon sources and future high-dimensional integrated quantum photonic systems.

\section{Methods}

\subsection{Whispering gallery mode selection and mesh convergence}

The whispering gallery mode spectrum was calculated using axisymmetric COMSOL eigenmode simulations. The horizontal (H)-like and vertical (V)-like mode families were identified by following each resonance as the mode order changed and by confirming their field distributions and confinement near the resonator rim. The nine selected resonances were then recalculated using progressively finer computational meshes to verify numerical convergence. The fine mesh was used for all reported results because further mesh refinement produced only negligible changes in the H--V resonance splitting compared with the loaded optical linewidth.

\subsection{Electro-optic tuning}

The electro-optic tuning was modeled by calculating the electric field produced by the lithium-niobate (LN) actuator and its effect on each whispering-gallery mode using the perturbation expression in Eq.~\eqref{eq:eo_perturbation}. The Pockels effect in the LN layer was used to determine how an applied voltage shifts the resonance frequency of each polarization mode. The operating voltage was then chosen to minimize the remaining frequency difference between the horizontal (H) and vertical (V) polarization modes across the nine selected resonances, expressed in units of the loaded optical linewidth.

\subsection{CMT/Jones reconstruction}

The relative strengths with which each frequency channel couples into and out of the resonator were calculated from the simulated optical fields using Eq.~\eqref{eq:relative_cmt}. These coupling strengths determine how each frequency channel transforms the photon's polarization and define the corresponding Jones matrices, including small polarization-mixing (leakage) terms. The Jones matrices were then applied to the signal photon in Eq.~\eqref{eq:source_state} to reconstruct the output polarization state after the frequency information was removed by tracing over the selected frequency channels. The reconstructed polarization state was then used to calculate the degree of entanglement (concurrence) and the maximum violation of the Bell inequality \(S_{\max}\) using standard methods \cite{Wootters1998,Horodecki1995, CHSH1969}.

\subsection{Frequency-bin state quality and photon count estimates}

The selected frequency channels were analyzed to determine how much information they can carry and how well the quantum coherence between them is preserved. We calculated the effective dimension \(K\), entropy \(H\), reduced frequency-bin purity, and fidelity to an ideal uniform nine-bin state from the normalized density matrix of the selected frequency channels.

We also estimated the photon detection and coincidence rates expected in an experiment. These estimates include the resonator coupling efficiency \(\eta_0\), the probability that a photon occupies one of the nine selected frequency channels (\(p_9\)), transmission through the optical components, detector efficiency, background counts, and accidental coincidences within the coincidence window.

\section{Conclusion}

We have presented a computational design for an electrically tunable microtoroid that selects the desired frequency channels from an externally generated pair of polarization-entangled photons while preserving their entanglement. The device is designed to operate across nine matched signal--idler frequency-channel pairs. A quarter-rim LN/electrode actuator provides a \(\SI{2.035}{\giga\hertz}\) polarization-dependent frequency correction at \(\SI{51.32}{\volt}\), reducing the remaining H/V frequency mismatch to just 0.286 loaded linewidths. Coupled-mode/Jones modeling predicts strong preservation of the polarization-entangled state, with \(C=0.969\), \(F=0.981\), and \(S_{\max}=2.785\). When the relative phases between the selected frequency channels are stabilized or programmed, the same device also supports a nine-channel polarization--frequency hyperentangled state with an effective dimension of \(K=8.97\) and an entropy of \(H=3.17\) bits.

The broader significance of this work is the demonstration that a compact, electrically tunable microresonator can connect conventional polarization-entangled photon sources with high-dimensional frequency-encoded quantum systems. Although the present study is computational, the design establishes clear performance targets for experimental realization, including accurate optical coupling, thermal and voltage stabilization, characterization of the source frequency distribution, and control of the relative phases between frequency channels. A practical proof-of-concept implementation could mount the thin-film lithium-niobate tuning element on a nanopositioner and position it near the microtoroid rim, analogous to tapered-fiber coupling in WGM experiments. This approach would allow the rim--film spacing to be adjusted experimentally while avoiding permanent contact with the high-\(Q\) resonator. Integrated fabrication of a fixed-gap LN tuning element would be a later engineering step. These results provide a practical foundation for developing microtoroid-based frequency selectors as compact interfaces for future integrated quantum photonic systems.

\section*{Funding}

National Institutes of Health (NIH) (R35GM137988); National Science Foundation (NSF) (2237077).

\section*{Disclosures}

J.S.: FemtoRays Technologies (I).

\section*{Data availability}

Data underlying the results presented in this paper are available from the corresponding author upon reasonable request.
\clearpage
\begin{center}
{\LARGE\bfseries Supplementary Information}\\[1ex]
{\large Design of an Electrically Tunable Microtoroid for Frequency Selection of Polarization-Entangled Photons}
\end{center}

\vspace{2em}

\section*{Supplementary Note 1: Selection of the D215 Microtoroid}

The D215 microtoroid was selected because it provided the best overall performance among the geometries examined for Bell-compatible frequency selection. The goal was to identify a resonator that could simultaneously select nine matched signal frequency channels while preserving the polarization entanglement of the photon pairs. Candidate geometries were therefore evaluated using both optical and quantum-state criteria, including the number of accepted frequency channels, the residual frequency mismatch between the horizontal- and vertical-polarized resonances after optimization with a single applied voltage, and the resulting Bell-state metrics.

The retained D215 design supports nine selected signal frequency channels and achieves a final horizontal--vertical resonance mismatch of only 0.286 loaded linewidths after electro-optic tuning with the quarter-rim lithium-niobate actuator. The corresponding quantum-state analysis predicts a concurrence of \(C=0.969\), a Bell-state fidelity of \(F=0.981\), a maximum CHSH Bell parameter of \(S_{\max}=2.785\), and an effective frequency-bin dimension of \(K=8.97\). Together, these results indicate that the resonator can select the desired frequency channels while maintaining strong polarization entanglement. Table~\ref{tab:s_d215_summary} summarizes the retained design used throughout the main manuscript.

\begin{table}[b]
\centering
\caption{Summary of the retained D215 microtoroid design used throughout the main manuscript.}
\label{tab:s_d215_summary}
\begin{tabular}{p{0.62\linewidth}c}
\toprule
Metric & Value \\
\midrule
Selected frequency channels & 9 \\
Toroid diameter & \(\SI{215.4}{\micro\meter}\) \\
Minor radius & \(\SI{4}{\micro\meter}\) \\
LN tuning voltage & \(\SI{51.32}{\volt}\) \\
Reference voltage & \(\SI{126.6}{\volt}\) \\
Residual H/V mismatch & 0.286 linewidths \\
Maximum H/V mismatch & 0.143 linewidths \\
Concurrence \(C\) & 0.969 \\
Bell-state fidelity \(F\) & 0.981 \\
Maximum CHSH parameter \(S_{\max}\) & 2.785 \\
Effective dimension \(K\) & 8.97 \\
\bottomrule
\end{tabular}
\end{table}
\section*{Supplementary Note 2: Mesh and Numerical Convergence}

To verify that the reported device performance is not limited by the numerical mesh, we compared four axisymmetric finite-element meshes: coarse, medium, fine, and extra-fine. The extra-fine mesh serves as the numerical reference, while the fine mesh is used throughout the main manuscript because it provides essentially the same results with substantially lower computational cost.

To isolate numerical discretization errors from variations associated with the electro-optic optimization, this convergence study uses a fixed differential tuning coefficient. Relative to the extra-fine mesh, the fine mesh changes the calculated resonance wavelength by at most \(\SI{0.031}{\pico\meter}\), the horizontal--vertical resonance splitting by at most \(\SI{16.1}{\mega\hertz}\), and the residual H/V mismatch by only \(0.0011\) loaded linewidths. These differences are much smaller than the linewidth scale relevant to the frequency-selection design. The final mode-dependent optimization presented in the main manuscript is performed separately using the fine mesh and gives an optimal tuning voltage of \(\SI{51.32}{\volt}\) with a residual H/V mismatch of 0.286 loaded linewidths.

\begin{figure}[t]
\centering
\includegraphics[width=0.86\textwidth]{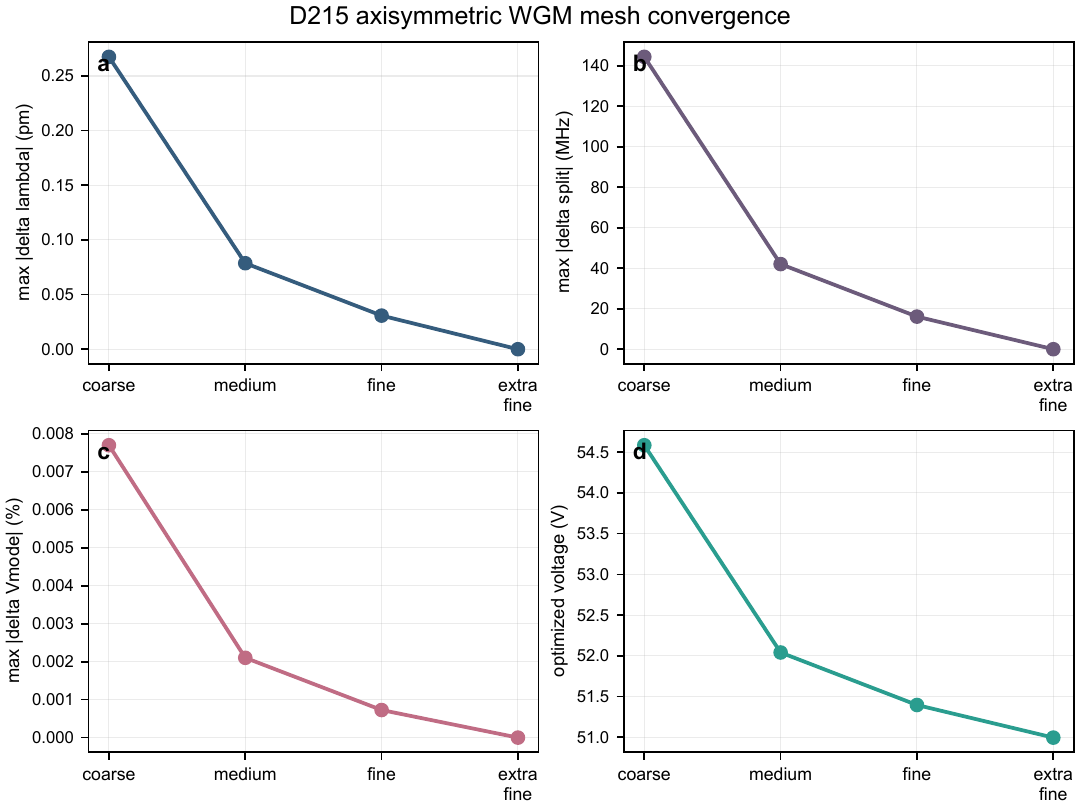}
\caption{Mesh-convergence study for the retained D215 microtoroid. Coarse, medium, fine, and extra-fine meshes are compared using the extra-fine mesh as the numerical reference. The fine mesh is used throughout the main manuscript because the remaining numerical differences are negligible compared with the optical linewidth of the resonator.}
\label{fig:s_mesh_convergence}
\end{figure}

\begin{table}[t]
\caption{Mesh-convergence results for the retained D215 microtoroid. All quantities are reported relative to the extra-fine mesh. The voltage and residual H/V mismatch are calculated using a fixed differential tuning coefficient to isolate mesh-related numerical errors. The final mode-dependent optimization reported in the main manuscript gives an operating voltage of \(\SI{51.32}{\volt}\) and a residual H/V mismatch of 0.286 loaded linewidths.}
\label{tab:s_mesh_convergence}
\centering
\small
\begin{tabular}{lcccccccc}
\toprule
Mesh &
Elem. &
Quality &
$\Delta\lambda$
(pm) &
$\Delta f_{\rm HV}$
(MHz) &
$\Delta V_m$
(\%) &
$\Delta F_s$
(p.p.) &
$V$
(V) &
Span
\\
\midrule
Coarse    & 43306  & 0.747 & 0.27 & 144.4 & 0.01 & 0.214 & 54.59 & 0.268 \\
Medium    & 76186  & 0.737 & 0.08 & 42.04 & 0.00 & 0.092 & 52.04 & 0.261 \\
Fine      &108710  & 0.745 & 0.03 & 16.11 & 0.00 & 0.046 & 51.40 & 0.259 \\
Extra fine&168738  & 0.728 & 0.00 &  0.00 & 0.00 & 0.000 & 50.99 & 0.258 \\
\bottomrule
\end{tabular}
\end{table}
\section*{Supplementary Note 3. Geometry Screening and Selection of the D215 Design}

The retained D215 microtoroid was selected from a broad computational survey of candidate geometries rather than from a single isolated simulation. Initial COMSOL parameter sweeps identified diameter ranges in which the horizontal- and vertical-polarized whispering-gallery-mode families could be brought into close alignment using a single applied voltage. These promising candidates were then examined in more detail by tracking each possible H-like and V-like mode family across the selected frequency channels and optimizing the single voltage for each candidate pair.

Candidate geometries were ranked according to three criteria: the number of accepted frequency channels, the residual horizontal--vertical resonance mismatch after single-voltage optimization, and the magnitude of the required electro-optic correction. This screening identified the D215 and D225 geometry families as the most promising candidates for Bell-compatible nine-channel frequency selection.

The final design reported in the main manuscript is based on the retained D215 geometry and uses the fine-mesh electro-optic and quantum-state analysis described in the main text. Tables~\ref{tab:s_microtoroid_focused_crossing} and~\ref{tab:s_microtoroid_broad_scan} summarize the geometry-screening results that led to the selection of the D215 design, while Supplementary Note~5 compares the retained microtoroid with representative microring geometries.
\begin{table}[t]
\centering
\small
\caption{Highest-ranked microtoroid branches identified during the focused geometry optimization. Candidate branches are ranked by the number of selected frequency channels, the required common H/V resonance correction, and the residual H/V resonance mismatch after optimization.}
\label{tab:s_microtoroid_focused_crossing}
\begin{tabular}{lcccc}
\toprule
Branch & Ch. & Corr. (GHz) & Span & Max. \\
\midrule
D215.4 (1308/1312) & 9 & 3.77 & 2.26 & 1.13 \\
D215.4 (1309/1313) & 9 & 3.88 & 2.22 & 1.11 \\
D215.4 (1313/1308) & 9 & 17.3 & 1.38 & 0.691 \\
D215.4 (1314/1309) & 9 & 17.3 & 1.36 & 0.679 \\
D215.4 (1310/1312) & 9 & 49.0 & 0.0434 & 0.0217 \\
\bottomrule
\end{tabular}
\end{table}

\begin{table}[t]
\centering
\small
\caption{Broad COMSOL geometry survey used to identify promising microtoroid diameter families for detailed optimization. The residual-error column was used only during the initial screening.}
\label{tab:s_microtoroid_broad_scan}
\begin{tabular}{lccc}
\toprule
Geometry & Ch. & Split (GHz) & Error (lw) \\
\midrule
D100 (elliptical) & 9 & 28.3 & 35.7 \\
D100 (circular) & 9 & \(1.35\times10^{3}\) & \(3.37\times10^{3}\) \\
D119.2 (circular) & 9 & 941 & \(2.30\times10^{3}\) \\
D215.4 (circular) & 9 & 1.35 & 1.69 \\
D225.1 (circular) & 9 & 2.15 & 2.68 \\
\bottomrule
\end{tabular}
\end{table}
\section*{Supplementary Note 4: Local Geometry Tolerance Study}

To evaluate the robustness of the retained D215 design, we performed a local parameter study around the nominal geometry (\(D=\SI{215.4}{\micro\meter}\), \(a=b=\SI{4.0}{\micro\meter}\)). Sixteen neighboring microtoroid geometries were recalculated using the same axisymmetric COMSOL model employed throughout this work. Candidate geometries were ranked using the same criteria as in the main manuscript: the number of selected frequency channels, the required common horizontal--vertical (H/V) resonance correction, and the residual H/V resonance mismatch after optimization.

Fifteen of the sixteen neighboring geometries retained all nine selected frequency channels, indicating that the design is not an isolated optimum. Two neighboring geometries, \(D=\SI{214.8}{\micro\meter},\,a=\SI{4.0}{\micro\meter}\) and \(D=\SI{215.0}{\micro\meter},\,a=\SI{4.0}{\micro\meter}\), produced slightly smaller residual H/V mismatches of 0.242 and 0.254 loaded linewidths, respectively. However, direct electro-optic simulations showed that these geometries couple much less efficiently to the present quarter-rim lithium-niobate tuning element, requiring approximately \(\SI{803}{\volt}\) and \(\SI{6.07}{\kilo\volt}\), respectively, to achieve resonance alignment. By comparison, the retained D215 design requires only \(\SI{51.32}{\volt}\) while maintaining a residual H/V mismatch of 0.286 loaded linewidths and preserving all nine selected frequency channels.

These results demonstrate that the retained D215 geometry represents the best overall trade-off between optical performance and practical electro-optic tunability, rather than simply the geometry with the smallest residual spectral mismatch. Table~\ref{tab:s_d215_local_tolerance} summarizes the local tolerance study. The reported \(S_{\max}\) values for neighboring geometries are screening estimates and were not obtained from the full coupled-mode/Jones-matrix analysis used for the retained design.
\begin{table}[t]
\centering
\small
\caption{Local geometry tolerance study around the retained D215 design. Candidate geometries are re-optimized using the same branch-selection procedure as in the main manuscript. The H/V correction is the common spectral correction obtained from the branch search. The \(S_{\max}\) values for neighboring geometries are screening estimates; the retained D215 value is obtained from the full coupled-mode/Jones-matrix analysis.}
\label{tab:s_d215_local_tolerance}
\begin{tabular}{ccccccc}
\toprule
\(D\) &
\(a\) &
Ch. &
H/V corr.
(GHz) &
Span &
Max. &
\(S_{\max}\) \\
(\si{\micro\meter}) &
(\si{\micro\meter}) &
&
&
(lw) &
(lw) &
estimate \\
\midrule
\IfFileExists{main_text_csv/d215_local_tolerance_table_rows_no_voltage.tex}{%
\input{main_text_csv/d215_local_tolerance_table_rows_no_voltage.tex}%
}{%
214.8 & 4.0 & 9 & 0.0849 & 0.242 & 0.121 & 2.822 \\
215.0 & 4.0 & 9 & 0.6350 & 0.254 & 0.127 & 2.821 \\
216.0 & 4.1 & 9 & 0.9980 & 0.824 & 0.412 & 2.759 \\
215.4 & 4.0 & 9 & 2.0350 & 0.286 & 0.143 & 2.785 \\
215.8 & 4.0 & 9 & 3.4900 & 0.286 & 0.143 & 2.820 \\
216.0 & 4.0 & 9 & 4.1900 & 0.299 & 0.149 & 2.819 \\
215.4 & 4.1 & 9 & 4.3500 & 0.854 & 0.427 & 2.755 \\
216.4 & 4.0 & 9 & 5.6200 & 0.308 & 0.154 & 2.818 \\
}
\bottomrule
\end{tabular}
\end{table}
\section*{Supplementary Note 5: Microring Control Study}

To provide a comparison with an alternative resonator geometry, we performed the same one-voltage optimization for a set of silica and silica--air microring resonators. Forty candidate designs were evaluated using the same nine-channel frequency-selection criterion applied to the retained D215 microtoroid. Each candidate was ranked according to the residual horizontal--vertical (H/V) resonance mismatch after optimization and the magnitude of the required common H/V resonance correction.

Several microring geometries produced very small residual H/V mismatches. However, these designs required unrealistically large resonance corrections to align the horizontal- and vertical-polarized mode families. The flattest branch achieved a residual mismatch of only 0.0174 loaded linewidths but required a common H/V correction of \(\SI{1.69}{\tera\hertz}\). Even the most favorable low-correction microring candidate still required \(\SI{475}{\giga\hertz}\), several orders of magnitude larger than the correction required for the retained D215 microtoroid.

\begin{figure}[t]
\centering
\includegraphics[width=0.92\textwidth]{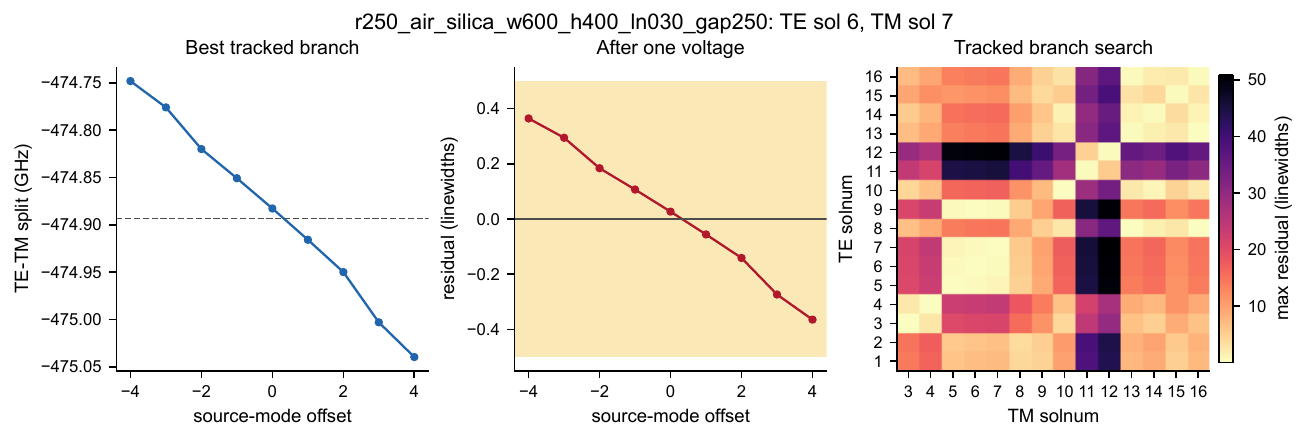}
\caption{Representative microring optimization result for the best tracked branch. Some microring geometries achieve smaller residual H/V resonance mismatches (Table~\ref{tab:s_microring_ranking}), but require large common H/V resonance corrections. This comparison shows that the residual mismatch alone is not sufficient for selecting a practical resonator geometry.}
\label{fig:s_microring_tracked_branch}
\end{figure}

Figure~\ref{fig:s_microring_tracked_branch} and Table~\ref{tab:s_microring_ranking} summarize the microring comparison. These results demonstrate that minimizing the residual H/V mismatch alone is insufficient; the required resonance correction must also be compatible with a practical electro-optic tuning mechanism.

\begin{figure}[t]
\centering
\includegraphics[width=0.92\textwidth]{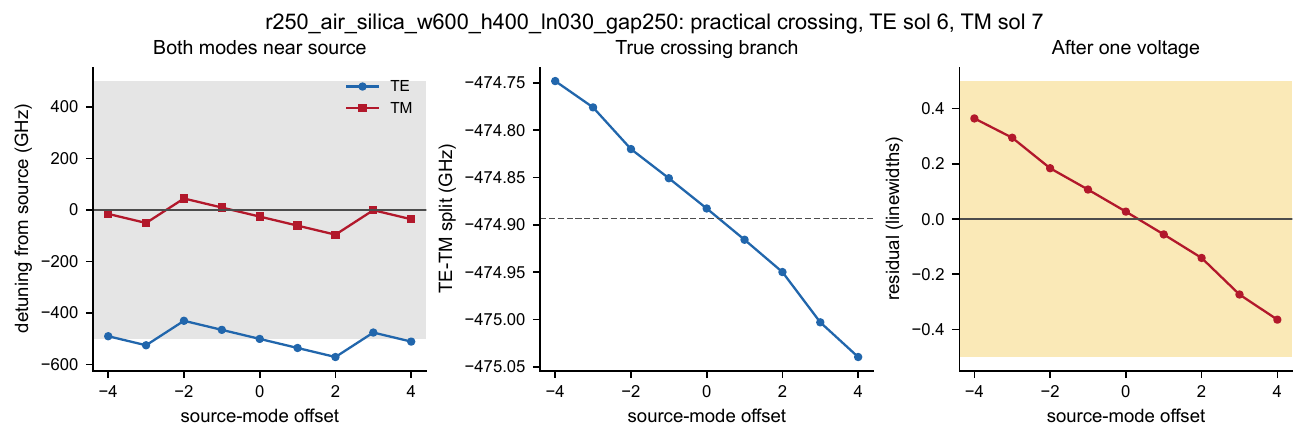}
\caption{Representative practical microring optimization result. The best practical microring candidate preserves all nine selected frequency channels and achieves a residual H/V resonance mismatch of 0.729 loaded linewidths, but still requires a common H/V resonance correction of \(\SI{475}{\giga\hertz}\). This comparison illustrates that both the residual mismatch and the required electro-optic tuning must be considered when evaluating candidate resonator geometries.}
\label{fig:s_microring_practical_crossing}
\end{figure}

\begin{table}[t]
\centering
\small
\caption{Highest-ranked microring branches from the expanded branch-crossing analysis. The first three rows are the best practical nine-channel candidates with the smallest required H/V resonance correction. The final row is the flattest branch, included for comparison. Residual spans are reported after the minimax common H/V correction.}
\label{tab:s_microring_ranking}
\begin{tabular}{lcccccc}
\toprule
Branch & Ch. & Corr. (GHz) & Span & Max. & Mean (GHz) & Mesh \\
\midrule
6/7   & 9 & 474.9  & 0.729  & 0.365  & 538.9  & 0.663 \\
10/14 & 9 & 522.4  & 1.870  & 0.934  & 4142.6 & 0.663 \\
5/7   & 9 & 525.3  & 1.790  & 0.896  & 589.3  & 0.663 \\
6/6   & 9 & 1692.4 & 0.0174 & 0.0087 & 2694.4 & 0.663 \\
\bottomrule
\end{tabular}
\end{table}
\section*{Supplementary Note 6: Scope of the Geometry Comparison}

The geometry studies presented in this Supplementary Information explain why the D215 microtoroid was selected for the main manuscript. Among the resonator geometries examined, it provided the best overall balance between the number of selected frequency channels, the residual horizontal--vertical (H/V) resonance mismatch after single-voltage optimization, and the magnitude of the required electro-optic tuning. The comparison is intended to identify the most suitable geometry for the present Bell-compatible, post-source frequency-selection architecture rather than to establish a universally optimal resonator design. Alternative resonator materials, geometries, or tuning mechanisms may achieve different trade-offs and could be optimized for other quantum photonic applications.

As discussed in the main manuscript, interpreting the device output as a polarization--frequency hyperentangled state additionally requires stable or actively controlled relative phases between the selected frequency channels. Without phase stabilization, the device functions as a Bell-compatible frequency selector that preserves polarization entanglement while selecting multiple frequency channels.

\section*{Supplementary Note 7: Additional Visible/NIR Wavelength Context}

The computational design presented in the main manuscript considers a source that generates polarization-entangled photon pairs at \(\SI{750}{\nano\meter}\) (signal) and \(\SI{880}{\nano\meter}\) (idler). These wavelengths lie within an active region of visible and near-infrared quantum photonics. Representative examples include WSe\(_2\) quantum-light emitters operating near \(\SI{750}{\nano\meter}\), diamond silicon-vacancy color centers near \(\SI{737}{\nano\meter}\), and GaAs quantum-dot sources of polarization-entangled photons near \(\SI{785}{\nano\meter}\) \cite{PalaciosBerraquero2016QLED,Rogers2014SiV,Sipahigil2016DiamondNetwork,BassoBasset2021QDQKD}. These examples illustrate that the wavelength range considered here is compatible with several established quantum photonic platforms.

\section*{Supplementary Note 8: Nonselected Resonances and Off-Bin Response}

The coupled-mode/Jones-matrix model presented in the main manuscript treats each of the nine selected resonance pairs as an isolated frequency channel. To verify this approximation, we examined the complete set of COMSOL eigenmodes for the retained D215 design after the branch-selection procedure. For each of the 18 selected horizontal- and vertical-polarized resonances, the nearest nonselected mode with the same azimuthal order was identified using the same fine mesh employed in the reported simulations.

The closest nonselected resonance is separated from the corresponding target resonance by \(\SI{16.50}{\giga\hertz}\), or approximately 41 loaded linewidths for \(Q_L=10^6\). Across all selected modes, the median nearest-neighbor separation is 162.7 loaded linewidths. At these separations, the Lorentzian response from neighboring resonances is negligible, supporting the approximation that each selected resonance can be treated as an isolated frequency channel.

We also examined all additional whispering-gallery-mode resonances lying within the \(\SI{750}{\nano\meter}\) signal wavelength range. These resonances are not part of the designed nine-channel frequency selector but could contribute background counts if a broadband photon source and broadband coincidence measurement were used. An approximate taper-coupling model was therefore used to estimate their accessibility.

Within the selected wavelength range, 73 off-bin WGM-like resonances were identified, of which 27 have estimated taper access comparable to or greater than the weakest selected channel. If all of these resonances were accepted by an unfiltered broadband coincidence measurement, the off-bin pair fraction would be \(f=3.53\) relative to the intended nine channels. Under the target-bin operating conditions considered in the main manuscript, however, prepared source bins together with wavelength-resolved idler recognition reject these additional resonances. Only the Lorentzian tails of the off-bin resonances contribute, reducing the off-bin fraction to \(f=3.62\times10^{-5}\), while maintaining \(S_{\max}=2.78514\) and \(C=0.96936\). These results support the isolated-channel approximation used throughout the quantum-state analysis.

\section*{Supplementary Note 9: Idler Frequency-Channel Identification}

The idler frequency channels corresponding to the retained D215 signal channels are determined from energy conservation,

\[
\lambda_{i,k}^{-1}
=
\lambda_p^{-1}
-
\lambda_{s,k}^{-1},
\]

using a pump wavelength of \(\lambda_p=\SI{405}{\nano\meter}\). Table~\ref{tab:s_idler_bins} lists the calculated idler wavelengths corresponding to the nine selected signal frequency channels used throughout the main manuscript. These wavelengths define the expected idler detection channels for coincidence measurements and frequency-bin identification.

\begin{table}[h]
\centering
\small
\caption{Energy-conserving idler wavelengths corresponding to the nine selected signal frequency channels. The mean idler-channel spacing is \(\SI{0.765}{\nano\meter}\) (\(\approx\SI{296}{\giga\hertz}\)). For comparison, the loaded linewidth at \(Q_L=10^6\) is approximately \(\SI{0.0010}{\nano\meter}\).}
\label{tab:s_idler_bins}
\begin{tabular}{cccc}
\toprule
Offset &
Signal
(nm) &
Idler
(nm) &
Next
(nm) \\
\midrule
\(-4\) & 751.8869 & 877.8486 & 0.7603 \\
\(-3\) & 751.3300 & 878.6090 & 0.7617 \\
\(-2\) & 750.7739 & 879.3706 & 0.7630 \\
\(-1\) & 750.2187 & 880.1336 & 0.7643 \\
\(0\)  & 749.6642 & 880.8979 & 0.7656 \\
\(1\)  & 749.1106 & 881.6636 & 0.7670 \\
\(2\)  & 748.5579 & 882.4305 & 0.7683 \\
\(3\)  & 748.0059 & 883.1988 & 0.7696 \\
\(4\)  & 747.4547 & 883.9684 & -- \\
\bottomrule
\end{tabular}
\end{table}
For a proof-of-principle experiment, the idler frequency channels can be measured by scanning a tunable Fabry--Perot filter across the nine predicted idler wavelengths using a single detector. In this architecture, the D215 microtoroid acts as the high-\(Q\) frequency selector for the signal photon, while the Fabry--Perot filter identifies the corresponding idler frequency channel through coincidence measurements. The Thorlabs FPQFA-8 provides a representative example, covering 845--1300~nm with a free spectral range of \(\SI{30}{\giga\hertz}\), finesse \(>300\), resolution \(<\SI{100}{\mega\hertz}\), and on-resonance transmission exceeding 80\% \cite{ThorlabsFPQFA8}. At \(\SI{880.9}{\nano\meter}\), this corresponds to an effective quality factor of \(Q_\mathrm{FP}>3.4\times10^6\).

The selected idler frequency channels have linewidths of approximately \(\SI{400}{\mega\hertz}\). The probability that an idler photon passes through the Fabry--Perot filter is therefore modeled as the overlap between the Lorentzian idler resonance and the Lorentzian Fabry--Perot transmission,

\[
T_\mathrm{bin}\simeq
T_\mathrm{peak}
\frac{\Delta\nu_\mathrm{FP}}
{\Delta\nu_\mathrm{bin}+\Delta\nu_\mathrm{FP}}.
\]

Using \(T_\mathrm{peak}=0.8\), the specified \(<\SI{100}{\mega\hertz}\) FPQFA-8 passband gives \(T_\mathrm{bin}\approx0.16\). Wider, passband-matched analyzers with \(Q_\mathrm{FP}=10^6\) and \(5\times10^5\) increase the predicted transmission to \(T_\mathrm{bin}=0.368\) and \(0.504\), respectively. These two cases are used in the source-rate and coincidence calculations presented in the main manuscript.

Because the Fabry--Perot free spectral range is approximately \(\SI{0.077}{\nano\meter}\) at \(\SI{880}{\nano\meter}\), multiple transmission orders are present. These additional orders do not produce valid coincidence events by themselves because each idler wavelength is correlated with a different signal wavelength through energy conservation. They can, however, increase the singles count and accidental coincidences. In practice, this effect can be reduced using coarse wavelength filtering, a larger-free-spectral-range Fabry--Perot, a grating-based demultiplexer, or a custom multi-channel filter. Similar approaches are widely used in frequency-bin quantum optics to identify correlated frequency channels \cite{Kues2017,Imany2018,Cheng2023SinglyFilteredBFC,Lu2023FrequencyBinReview}.

The source-rate estimates reported in the main manuscript are expressed in terms of the selected photon-pair rate after the D215 signal selector and the idler frequency analyzer, but before detector losses. For a prepared nine-channel source, the required source rate is determined by the signal-selector transmission and the finite Fabry--Perot overlap. For a broadband source, an additional factor accounts for the fraction of photon pairs that fall within the nine selected signal channels. For a uniform spectrum spanning the selected D215 signal window, this fraction is \(p_9=1.52\times10^{-3}\), corresponding to a \(657\times\) increase in the required source brightness before accounting for the Fabry--Perot transmission. For a uniform \(\SI{10}{\nano\meter}\) source bandwidth, the corresponding fraction is \(p_9=6.75\times10^{-4}\), requiring a \(1482\times\) increase in source brightness.

The complete parameter sweeps used to generate the source-rate and coincidence calculations are provided in the accompanying CSV files and supporting analysis scripts.

\section*{Supplementary Note 10: Scaling to Larger Frequency-Channel Windows}

The main manuscript considers a nine-channel frequency selector defined by the measured \(\SI{10}{\nano\meter}\)-wide signal window. This Supplementary Note examines how the same D215 microtoroid design would perform if a broader photon source and wider spectral window were available. These calculations are intended as a design study and should not be interpreted as predictions for the present experimental implementation.

To isolate the performance of the resonator itself, all selected frequency channels are assigned equal weights in this analysis. The calculations use the same optical modes, electrostatic simulations, electro-optic perturbation theory, and coupled-mode/Jones-matrix model employed throughout the main manuscript. The \(N=9\) case therefore serves as an equal-weight reference and differs from the measured-filter result presented in the main text, where the nonuniform source spectrum gives an effective frequency-bin dimension of \(K=8.97\) and a Shannon entropy of \(H=3.17\) bits. The equal-weight \(N=9\) design requires an operating voltage of \(\SI{51.32}{\volt}\), has a residual horizontal--vertical resonance mismatch of 0.286 loaded linewidths, and a maximum residual mismatch of 0.143 loaded linewidths.

The simulations indicate that the D215 geometry maintains strong performance as the number of selected frequency channels increases. High Bell-state fidelity and concurrence are maintained through approximately \(N=21\), while the maximum CHSH Bell parameter remains above the classical limit through \(N=27\). Beyond \(N=21\), the gradual increase in residual polarization imbalance reduces the quantum-state quality, although Bell-inequality violation is still predicted.

Broadband polarization-entangled photon sources provide useful context for these larger frequency windows. Jabir and Samanta demonstrated a temperature-tunable Type-0 PPKTP source pumped at \(\SI{405}{\nano\meter}\), reaching signal and idler wavelengths near \(\SI{743}{\nano\meter}\) and \(\SI{891}{\nano\meter}\), respectively, with a \(\SI{31}{\nano\meter}\) bandwidth at degeneracy \cite{Jabir2017HighBrightness}. Broadband Type-0 PPKTP sources have also demonstrated polarization visibilities of \(97.7\%\) over approximately \(\SI{100}{\nano\meter}\) \cite{Lohrmann2020BroadbandPPKTP}, while chirped Type-II PPKTP sources have achieved polarization-entangled bandwidths approaching \(\SI{125}{\nano\meter}\) \cite{Fraine2012BroadbandPPKTP}. These demonstrations suggest that broader frequency-channel windows are experimentally feasible, although any specific implementation should use the measured source spectrum or joint spectral amplitude to determine the actual channel weights.

\begin{figure}[t]
\centering
\includegraphics[width=\textwidth]{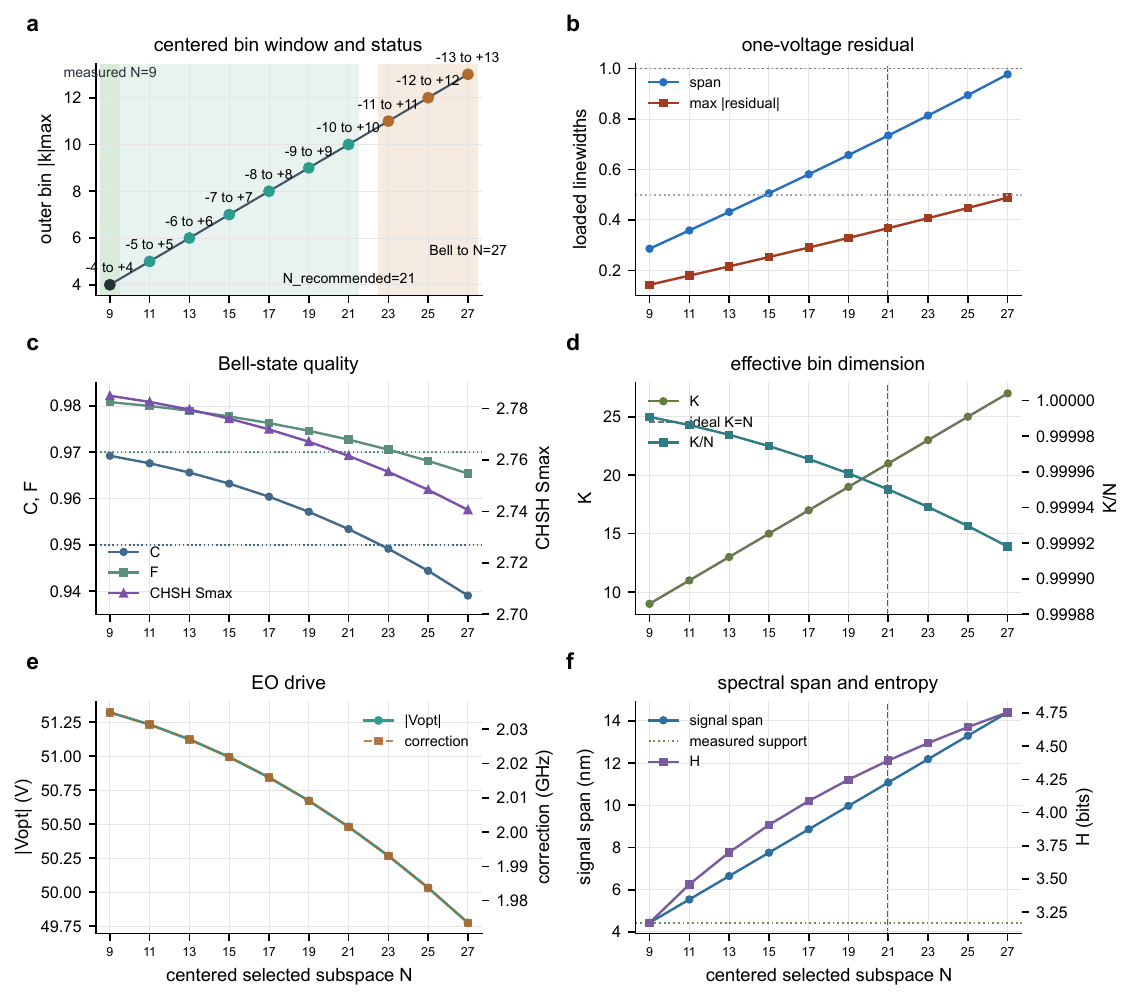}
\caption{Predicted performance of the D215 microtoroid as the number of selected frequency channels is increased. (a) Centered frequency-channel windows from \(N=9\) to \(N=27\). (b) Residual horizontal--vertical resonance mismatch after single-voltage optimization. (c) Predicted Bell-state metrics, showing that the maximum CHSH Bell parameter remains above the classical limit through \(N=27\). (d) Effective frequency-bin dimension \(K\) and channel utilization \(K/N\). (e) Optimized operating voltage and common H/V resonance correction. (f) Signal spectral span and Shannon entropy. Results for \(N>9\) assume a broader photon source and wider spectral window than those considered in the main manuscript.}
\label{fig:s_wide_source_filter_scaling}
\end{figure}
\begin{table*}[t]
\centering
\small
\caption{Predicted performance of the D215 microtoroid as the number of selected frequency channels is increased. All calculations assume equal channel weights. The \(N=9\) row is the equal-weight reference; the measured-filter result reported in the main manuscript gives \(K=8.97\) and \(H=3.17\) bits. Residuals are reported in loaded-linewidth units.}
\label{tab:s_wide_source_filter_scaling}

\begin{tabular}{cccccccccc}
\toprule
\(N\) &
Signal span (nm) &
Residual span &
\(V_{\rm opt}\) (V) &
\(C\) &
\(F\) &
\(S_{\max}\) &
\(K\) &
\(H\) (bits) &
Status \\
\midrule
9  & 747.46--751.89 & 0.286 & 51.32 & 0.969 & 0.981 & 2.785 & 9.00  & 3.170 & High quality \\
11 & 746.90--752.45 & 0.358 & 51.23 & 0.968 & 0.980 & 2.783 & 11.00 & 3.459 & High quality \\
13 & 746.36--753.00 & 0.432 & 51.12 & 0.966 & 0.979 & 2.780 & 13.00 & 3.700 & High quality \\
15 & 745.81--753.56 & 0.506 & 50.99 & 0.963 & 0.978 & 2.776 & 15.00 & 3.907 & High quality \\
17 & 745.26--754.12 & 0.581 & 50.84 & 0.960 & 0.976 & 2.772 & 17.00 & 4.087 & High quality \\
19 & 744.71--754.68 & 0.657 & 50.67 & 0.957 & 0.975 & 2.767 & 19.00 & 4.248 & High quality \\
21 & 744.17--755.25 & 0.735 & 50.48 & 0.953 & 0.973 & 2.762 & 21.00 & 4.392 & High quality \\
23 & 743.62--755.81 & 0.814 & 50.27 & 0.949 & 0.971 & 2.755 & 23.00 & 4.524 & Bell qualified \\
25 & 743.08--756.37 & 0.895 & 50.03 & 0.944 & 0.968 & 2.748 & 25.00 & 4.644 & Bell qualified \\
27 & 742.53--756.94 & 0.977 & 49.77 & 0.939 & 0.965 & 2.741 & 27.00 & 4.755 & Bell qualified \\
\bottomrule
\end{tabular}
\end{table*}

The measured-filter implementation remains the nine-channel design presented in the main manuscript. Among the broader-window designs examined here, the \(N=21\) configuration represents the most promising next step because it maintains the high-quality performance criteria while extending the operating range to \(\lambda_s=\SIrange{744.165}{755.246}{\nano\meter}\) and \(\lambda_i=\SIrange{873.314}{888.614}{\nano\meter}\). Although the \(N=27\) design continues to violate the CHSH Bell inequality, its concurrence and Bell-state fidelity no longer satisfy the more stringent quality criteria adopted in this work. Future studies of the \(N=21\) design should incorporate the measured source spectrum or joint spectral amplitude, extend the off-bin WGM analysis to the larger spectral window, and evaluate fabrication tolerances over the expanded operating range.

\section*{Supplementary Data and Reproducibility}

The accompanying supplementary data include the processed datasets and analysis scripts used to generate the results presented in this Supplementary Information. These files reproduce the mesh-convergence study, local geometry-tolerance analysis, microtoroid geometry screening, microring comparison, off-bin resonance analysis, idler frequency-selection calculations, source-rate estimates, and broader frequency-channel scaling studies. The normalized geometry-selection and quantum-state metrics are determined from the electromagnetic simulations and coupled-mode/Jones-matrix analysis and do not depend on an assumed absolute coupling efficiency. Larger COMSOL field datasets and additional exploratory simulations are retained separately from the processed summary files.

\bibliography{references_main, judy4}

\end{document}